# Sequence reproduction, single trial learning, and mimicry based on a mammalian-like distributed code for time

J. J. Hopfield[1] and Carlos D. Brody[1,2]
[1]Princeton Neuroscience Institute and Department of Molecular Biology
[2] Howard Hughes Medical Institute
Princeton University, Princeton NJ 08544-1014

**Abstract**

Animals learn tasks requiring a sequence of actions over time. Waiting a given time before taking an action is a simple example. Mimicry is a complex example—e.g. in humans, humming a brief tune you have just heard. Re-experiencing a sensory pattern mentally must involve reproducing a sequence of neural activities over time. In mammals, neurons in prefrontal cortex have time-dependent firing rates that vary smoothly and slowly in a stereotyped fashion. We show through modeling that a 'Many are Equal' computation can use such slowly-varying activities to identify each timepoint in a sequence by the population pattern of activity at the timepoint. The MAE operation implemented here is facilitated by a common inhibitory conductivity due to a theta rhythm. Sequences of analog values of discrete events, exemplified by a brief tune having notes of different durations and intensities, can be learned in a single trial through STDP. An action sequence can be 'played back' sped up, slowed down, or reversed by modulating the system that generates the slowly changing stereotyped activities. Synaptic adaptation and cellular post-hyperpolarization rebound contribute to robustness. An ability to mimic a sequence only seconds after observing it requires the STDP to be effective within seconds.

**Introduction**

An understanding of the passage of time is displayed in many short term phenomena at the psychological level. Trained to remember the time interval between a tone onset and reward availability, a rat can do so with an accuracy of ~25% (Catania, 1970; Roberts, 1981). Given a single instance of an irregularly spaced set of five clicks spread over two seconds, a person can tap out the pattern with good fidelity. Any musician can hear a single instance of a brief melody and replay it with fidelity. All of us experience this short-term 'tape recorder in the head' phenomenon in which we hear somebody to speak, briefly remembering not just the recent phrase, but also the tone and voice in which it was said.

Extracellular recordings in rats and monkeys doing tasks involving time intervals lasting seconds have revealed neurons that have rates of action potential firing that change slowly during the time interval (Brody et al., 2003; Kojima and Goldman-Rakic, 1982; Komura et al., 2001; Mita et al., 2009). An example, taken from prefrontal cortex of monkeys, is shown in Fig. 1a. Each neuron has a different but stereotyped activity pattern. The time since the onset of the trial can be decoded from the pattern of activities across neurons in prefrontal cortex, a large fraction of which are active at any particular time. The situation may be contrasted with the representation of time that has been found in area HVC of songbirds (Hahnloser et al., 2002). During singing, individual HVC cells fire a short burst of action potentials at a single timepoint during a syllable, thus explicitly encoding that timepoint in a labeled-line code. That is, in songbirds, asking "what time is it now?" can be easily answered by asking "which cell is firing now?" In contrast, when a substantial fraction of cells are firing, as in the mammalian prefrontal cortex example, it is the activity pattern over the population of cells with time-dependent cell firing rates that encodes the timepoint. If this slowly-changing, distributed coding is the fundamental representation of time in mammals, how can times of events be encoded, learned, and decoded in order to generate sequences? The encoding and learning must be flexible enough to recapitulate a sequence of events and intensities.

Higher animals have the capacity to reproduce a sequence of observed actions, as in mimicry. The ability to regenerate, in some appropriate part of the brain, the same spatiotemporal pattern of neural activity as was produced by the initial stimulus after an observation of the pattern, is a plausible neural basis for mimicry. Often, only a single observation of the sequence is required in order to regenerate it. We show that spike timing dependent plasticity of an appropriate form can result in single trial learning of such sequences

In this paper we describe a cell and circuit model which reproduces this class of phenomena.  The approach is appropriate for both sequences of multiple time intervals and for single time intervals (Matell and Meck 2004; Meck, 2005; Mauk and Buonomano, 2004) that involve only knowing that a given amount of time (since a starting point) has elapsed.

The problem is more complex for a prefrontal-like distributed representation of time than they are for HVC-like labeled-line timecodes. Our focus here is on the distributed system.  We show that a common rhythm to an array of cells whose overall firing rates are like those shown in Fig. 1, combined with synapse spike-timing-dependent plasticity, provides a solution to these computational problems of sequence production and learning.  We will construct a neural system that embodies a map of (timepoints)-->(internal recapitulation or motor acts). The central idea is that any particular time can be recognized by its pattern of activation across a set of neuronal activities by an implementation of the Many Are Equal (MAE) algorithm   (Hopfield and Brody, 2001; Hopfield and Brody, 2000), made possible by the spike timing effects of a common rhythm  (Brody and Hopfield, 2003).

Although the basic idea requires only very simple neurons and synapses, synaptic and cellular adaptation are useful in achieving operation across a range of conditions.  Unlike some schemes for learning analog quantities such as intensities (Brozović et al., 2007; Zipser and Andersen, 1988). the encoding of the learned analog variables is not encoded in synaptic strengths *per se*, but rather in the selection of which connections to be made.  Synapse modification is thus a yes-no question, not a how-much question, and this fact permits rapid learning of analog values.

In previous modeling work, we have implemented MAE by three different means, namely using a common gamma oscillation (~40 Hz) as input to a group of neurons   (Brody and Hopfield, 2003); similarly using a noisy gamma oscillation   (Markowitz et al., 2008); or through direct synaptic connections   (Hopfield and Brody, 2001). We introduce here a fourth implementation of MAE using a common theta rhythm (~8 Hz) and conductance-based inhibition. This method is appropriate if focusing on slower dynamics (unlike   (Hopfield and Brody, 2001; Markowitz et al., 2008)) and permits the learning of intensity-dependent information (unlike the intensity-independent representation sought in (Brody and Hopfield, 2003)).

The network to be simulated and simulation results are described in the bulk of the paper.  The concluding section sets the results and meaning of these simulations in the experimental context of animal systems and behaviors, and expected concomitants at the neuronal level.

**Results**

*The model's conceptual framework*

Time-dependent activity of the type shown in Fig. 1a can be used as an implicit code for elapsed time. Different neurons are found to have very different, but highly reproducible, temporal signatures (Brody et al., 2003). This allows identifying specific timepoints by identifying a specific set of neurons whose firing rates are all similar at the desired timepoint. The confluence of the firing rates of those neurons is then the signature of that timepoint (Fig. 1b, colored circles). Furthermore, at any given timepoint we can choose confluences at different firing rates to indicate different possible intensities (Fig. 1b, blue and green circles). In previous modeling work, we have called detection of such confluences, that is, firing rate similarity across a pool of neurons, a "Many Are Equals" computation. We have also shown how this computation can be implemented neurally: a shared oscillation can be used to induce spike synchrony across the chosen pool of neurons when and only when their firing rates are similar   (Hopfield and Brody, 2001).  It is useful to think of the problem of representing the motor acts necessary to play a simple melody, representing a sequence of notes at appropriate times and intensities.

Fig. 1d shows the overall architecture of the model we shall develop here. Prefrontal-like neurons, encoding time through slowly varying firing rates and called here "P" neurons, excite a set of neurons that we call the "S" neurons. The S neurons also receive input from a common signal that serves to synchronize the spike times of those S neurons that have similar input currents. A different set of neurons, the γ neurons, explicitly represent, in one-to-one fashion, the set of possible notes to be played (or, more generally, the set of motor acts to be carried out). A given note at a given time is then represented by creating connections from those S neurons that have confluent inputs at the chosen time to the γ neuron that corresponds to the chosen note. The intensity of the note to be played is encoded by the firing rate at which the confluence of the selected S neurons occurs.

How are the stereotyped P neuron activities generated? There are many neurally plausible answers to this question, and for the purposes of this paper it does not much matter which is used. One convenient possibility, used in the simulations here, is to have a network of neurons whose state of activity moves slowly along a collective coordinate, and which controls in a coordinated fashion the activities of all the P neurons (see Methods). This results in approximately Gaussian-shaped, slowly changing patterns of activity, as shown in Fig. 1b.

When time intervals involved in a known behavioral task are sped up or slowed down, it has been observed experimentally that firing patterns of the type of Fig. 1a are correspondingly compressed or stretched in time (Brody et al., 2003; Komura et al., 2001) Here, speed of P neuron replay corresponds to the speed of motion along their collective coordinate. If this is fast (slow), the event map is read out rapidly (slowly), and the internal 'melody' is played rapidly (slowly). Rapid replay of sequences previously observed through experience, or reverse replay, as in (Lee and Wilson, 2002; Foster and Wilson, 2006; Diba and Buzsáki, 2007), can thus be obtained simply by increasing the speed of motion along the collective coordinate of the slowly varying firing rates (Fig. 1c), or by reversing its direction of motion.

A cornerstone of the above model architecture is the detection of confluences, the "Many Are Equals" (MAE) computation. Given a set of variables $P_k$, the MAE algorithm answers the question of whether a substantial subset (the "Many") of the $P_k$ are approximately equal. The algorithm is significant to neurobiology because it can be readily implemented by spiking neurons when the variables $P_k$ are the input currents to a set of neurons. Synchronization of the neurons having equal values of input currents can occur through a variety of mechanisms: when the neurons are synaptically interconnected (Hopfield and Brody, 2001), or when there is a common periodic gamma oscillatory input to the set of neurons (Brody and Hopfield, 2003), or when there is a common noisy gamma oscillatory signal incident on the set (Markowitz et al., 2008). This synchronization (and its absence when the $P_k$ are not approximately equal) is the neurodynamic basis for the MAE operation.

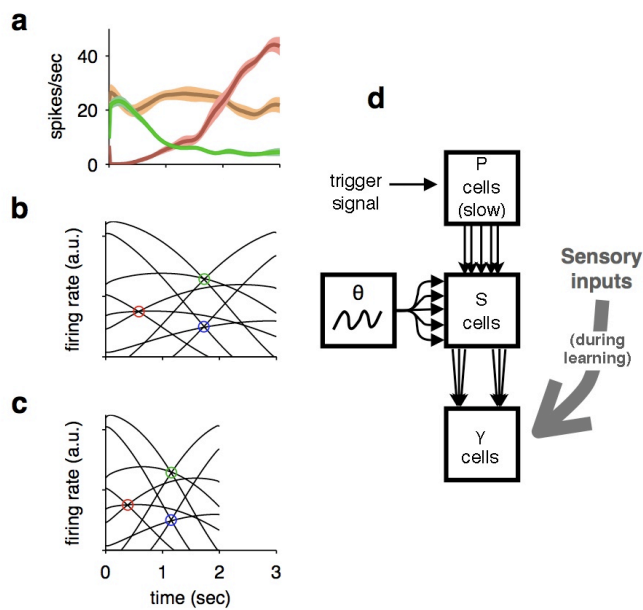

**Figure 1.** Firing rates that reliably change on a timescale of seconds to tens of seconds can be used to represent temporal intervals. a) An example of slow time signatures in the firing rates of 3 different neurons, recorded from prefrontal cortex of monkeys performing a sequential discrimination task. Data for this figure was kindly provided by R. Romo. All data in this panel is from identically prepared trials, recorded from a single monkey. Each neuron's firing rate is shown in a different color. The solid line represents the average of 20 trials; shading indicates the standard error of this average. Time signatures can be seen to be reliable, and very different across different neurons. b) The slowly changing firing rates used in the current modeling study. Each black line represents a different neuron. Neurons are drawn from a standard model of a slowly moving "bump" attractor (Wang, 2001). Any given timepoint can be identified by choosing a set of neurons whose firing rates are all similar at that time point (colored circles), a "Many Are Equals" computation (Hopfield and Brody, 2001). In addition, at any given time point different intensities can be identified by choosing different sets of confluences (e.g. green and blue circles). c) Connectivity between different components of the model used here. Neurons with slowly changing firing rates and a common noisy oscillation both impinge on a set of neurons, labeled S. Synchronous action potentials in a group of S neurons indicates that the input drives to this group are all similar, which is the signature of a particular timepoint. Integrate-and-fire neurons with a high firing threshold, acting as coincidence detectors, report the output of this "Many Are Equals" computation (γ neurons). In addition, if the γ neurons are driven by sensory inputs at the appropriate time, a spike-timing-dependent learning rule can be used to learn the S-to-γ connections. d) The speed of the spatiotemporal sequence playback can be accelerated or decelerated simply by changing the speed of the slowly moving bump attractor. A 1.5x speedup is illustrated.

Given the slowly-changing P neuron firing patterns, then, subsets of S neurons of Fig. 1d can use MAE to detect chosen timepoints through spike synchrony.

*Theta-based "many are equals"*

In this work, we have explored and implemented the MAE operation based on a common theta signal that activates an inhibitory conductance. This mechanism turns out to achieve rapid synchronization over a wider range of firing rates than previously described MAE mechanisms when the inputs vary on a timescale of hundreds of milliseconds or slower. The capacity to encode intensity rapidly over a wide range is important for the type of temporal sequences that are the focus of this paper.

Theta-based synchrony can be understood by first considering an extreme case in which an intuitive answer is exactly correct. We examine a set of spiking leaky integrate-and-fire neurons, each receiving the same steady input current, and each with its own small membrane noise. The membrane potential as a function of time for a set of 5 such neurons is shown in Fig. 2a. The threshold for firing is at 10 mV, and when the membrane potential reaches that level, an action potential is generated and the membrane potential switches to the 'reset' level. Prior to t = 1, the membrane potentials and action potentials of the different neurons are uncorrelated. At t = 1, a strong brief inhibitory conductivity pulse (*e.g* for $K^+$ or $Cl^-$ ions) is introduced to all neurons. If this conductivity pulse is large enough, it will result (as in the figure) in the membrane potentials of all the neurons being driven to the Nernst potential for that ion, in this figure taken as -0.002 V. When this pulse ends at t = 1.005, the neurons resume their firing. However, because they now all have the same membrane potential, the trajectories of membrane potential for all the neurons will be very similar, and they will fire almost synchronously for a while until the noise in the system decorrelates them, a behavior clearly visible in the action potential bunching after t = 1 sec at the top of Fig 2a.

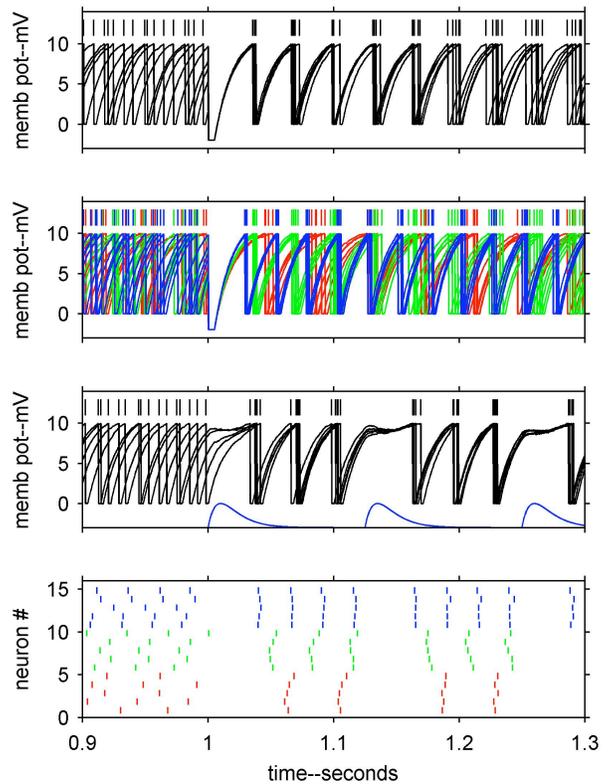

**Figure 2** a) membrane potential (mV) vs time for 5 spiking model neurons receiving the same input current. Threshold for spike generation 10 mV, reset potential 0 mV. Spikes are indicated above. At t=1.0, a large 5msec duration conductivity pulse occurs, briefly clamping the cell potential at the Nernst potential for an inhibitory ion, here taken as -2 mV. b) The same as a) except that three groups of 5 cells are shown, with different excitatory input

The time at which the first synchronous event occurs is determined by the time it takes the membrane potential to change from the Nernst potential at which it was briefly clamped to the firing threshold. This time depends on the steady excitatory current flowing into the cell, and is approximately 1/(firing rate) of the neuron. Thus if the 5 neurons had very different excitatory input currents and intrinsic firing rates, the first action potentials upon release from inhibition would occur at very different times, and there would be no highly synchronous event near t = 1.05 sec.  Fig. 2b depicts the same kind of behavior, except now for 3 groups of 5 neurons.  Neurons within each group all receive the same excitatory current, but the excitatory currents for the different groups are different.  Synchrony within each group after the inhibitory pulse is evident, and because they have different excitatory currents, different groups synchronize at different times.  A single large inhibitory conductance pulse can thus implement the MAE operation, synchronizing groups of neurons where Many have approximately Equal driving currents. In addition, since the separation between the pulses of synchronized action potentials is the firing rate of the neurons due to their excitatory input currents, the separations between the synchronized action potential clumps of each color encode the input current level for the color.

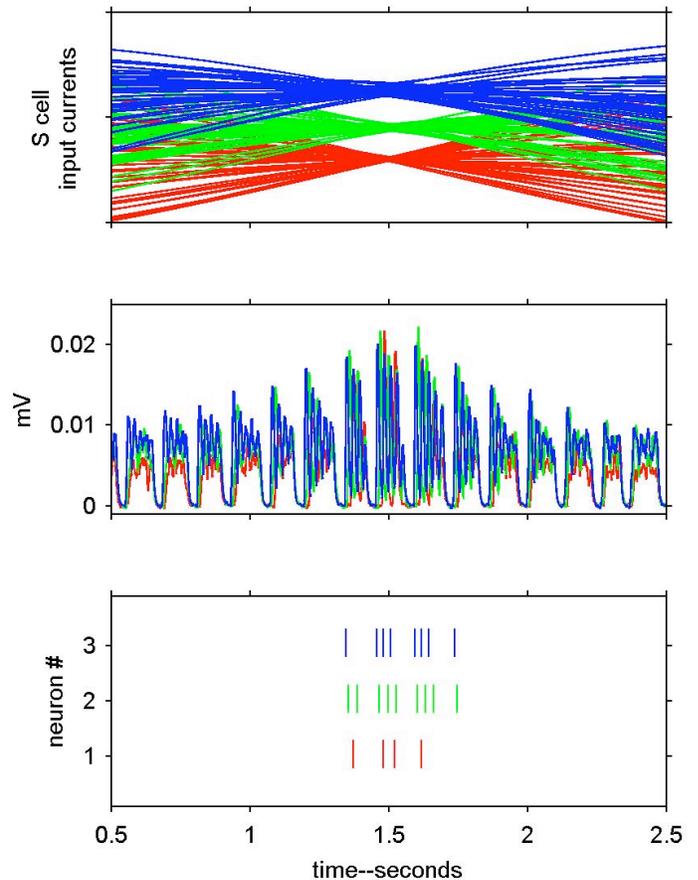

Fig 2c shows that the basic phenomenon behaves similarly in a less idealized situation, when the inhibitory conductivity pulses are modest in size and have an extended time course like that which would occur at a $GABA_B$ synapse responding to an action potential (Levitan and Kaczmarek, 1996)  This figure is like Fig 2a except that there are three extended conductivity pulses occurring at a rate of 8 Hz, whose time dependence is shown at the bottom of Fig 2c.  The size of the peak potassium conductivity represented by these pulses corresponds to a reduction of the membrane time constant by 20% in these simulations.   Fig 2d corresponds to Fig 2b in showing the action potentials of three groups of five neurons synchronized by the conductivity pulses shown in Fig 2c.  Once again the action potentials of different neurons receiving the same excitatory current become approximately synchronized by the inhibitory conductivity pulses.  In this case, the synchronous pulses come in groups, and the separation between the action potential clumps within a group implicitly encodes the input current to the group. Thus the presence of synchronous events reports that many inputs are equal; and the frequency of the synchronous events reports the level at which the inputs are equal.

*Playing a melody*

Playing a melody involves producing notes of specific intensities at defined times. Let us take the specific example of playing a brief chord composed of three notes of duration 0.3 sec, each at a different intensity, 1.5 s. after the beginning of the melody. We will assume that intensity corresponds to γ neuron firing rate. In the theta-based MAE mechanism we have described above, a desired γ neuron firing rate corresponds to a particular input current level to the S neurons. In the examples below, we will use input current levels of 0.3, 0.45, and 0.6 to represent three different intensities. For each current level, a subset of 30 S neurons are identified, which all have very nearly the target input current at the chosen time (1.5 sec).

**Figure 3** a) Input currents around t=1.5 s for 30 selected S neurons. Each line corresponds to the input from one P neuron; only inputs that at t=1.5 are close to a current of 0.3 (red), 0.45 (green), and 0.6 (blue) are shown. Finding neurons that have currents that are very similar to our targets at t=1.5, but diverge elsewhere, is possible thanks to the large diversity of S neurons available. b)  The membrane potentials of the three designed γ neurons that each receive synapses from one of the selected groups of S cells. In this panel, the γ neurons have been prevented from spiking so as to best illustrate the membrane potential.  c) The spikes from the three designed γ neurons using adapting synapses.
5

The time dependencies of the input currents of these three sets of neurons is shown in Fig 3a. Each of these sets of neurons is connected to its own recognition cell (γ neuron) by fast excitatory synapses. The recognition cells have been taken to have membrane time constants of 5 msec. For this particular example, we chose three simultaneous notes, but different times could equally well have been chosen for each note.

Fig. 3b shows the membrane potential of the recognition cells when spiking in these cells is suppressed. The y-axis units are arbitrary since they scale with the synaptic strength chosen. The S cells for each of the three γ cells become synchronized at approx 1.5 sec, due to the MAE mechanism, leading to a substantial set of high-magnitude peaks in the γ cells' membrane potential near t = 1.5 sec. For any of the three cells, a threshold can be chosen such that the γ cell fires chiefly during the desired interval. However, the threshold necessary to do this is different for the three cases: any single choice would lead to a longer duration for the higher-intensity note. The differences are due to the fact that the precision of synchrony in the presence of noise depends on the size of the input current. Because smaller input currents lead to larger time intervals or time delays, there is more time for the effects of noise to produce decorrelation. In order to make a system that will work gracefully over a range of input currents, an adaptive mechanism is needed. We have chosen to use synaptic adaptation, and to have a single non-adapting threshold level for all γ cells.

The effectiveness of many kinds of synapses is reduced by recent presynaptic neuronal firing (as for example by vesicle depletion), producing a firing-rate-dependent synaptic efficacy (Tsodyks and Markram, 1997; Abbott et al., 1997). Such adaptation has been included in the modeling (see Methods). With this inclusion, all excitatory synapses to γ cells have the same properties, all γ cells have the same properties, and the system functions well over a range of input currents.

Figure 3c shows the spikes of the three γ cells, when synaptic adaptation is included is S to γ cell synapses, and the γ cells are allowed to spike. All three γ cells have the same membrane potential firing threshold. At 1.5 sec., these cells all fire a multiplet after a theta-inhibitory event. The spike intervals in this multiplet encode the size of the analog current that was selected by each recognition cell.

The same ideas can be used to design the appropriate connections for a network that can recapitulate a 10 element sequence having a mixture of event durations and intensities. This is shown in Fig. 4.

We have seen that accurate intensity information is contained in the spike intervals of multiplets. For some purposes, the mere existence of such a spike interval representation is all that may be necessary. For other neurobiological purposes (e.g. driving a muscle) it may be desirable to 'neurally' decode a spike train into an analog pattern. The decoder should be sensitive to the interspike interval between pairs of γ cell spikes: the shorter this is, the greater the intensity. One straightforward system sensitive to interspike intervals is an integrate-and-fire neuron, because the maximum membrane potential observed in response to two incoming spikes is sensitive to the temporal separation between the spikes. In the Methods section, we describe analog decoders based on using γ cell spikes to drive a population of integrate-and-fire neurons, thus obtaining a spike-interval-sensitive readout, and averaging over this population of readout neurons, thus obtaining an analog signal. Applied to the spike trains of Fig. 4b, it produces the results illustrated in 4c and 4d. While this decode does not accurately recapitulate the uniformity of intensity within each element of the sequence of the top panel, its peaks do a fine job of encoding the analog values. When this reconstruction is used to control a tone generator, and the sequence has been chosen to represent a simple melody, the neural playback even from using a single γ cell for each note is strikingly similar to the original melody.

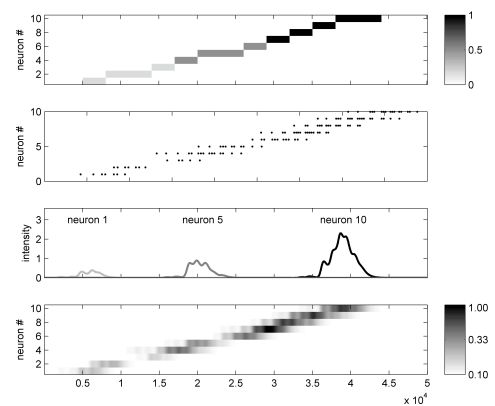

**Figure 4** Appropriate choice of S to γ cell connections can produce a desired sequence of γ cell firing rates. a) The desired intensities (gray scale) vs time for a sequence of 10 elements with varying intensities and durations. Three different levels of intensity and 2 different durations are present in the sequence. b) The action potentials of 10 γ neurons from a network with S to cell connections designed to produce the sequence shown in (a). The intervals in the γ cell action potential bursts encode the corresponding intensities (see Fig. 3). c) Analog intensity vs time can be decoded using a simple 'neural decoder'; the decoder's output is shown for γ neuron numbers 1,5, and 10. d) Same analog intensity 'neural decode' as in (c), but shown using gray scale for all 10 γ cells.

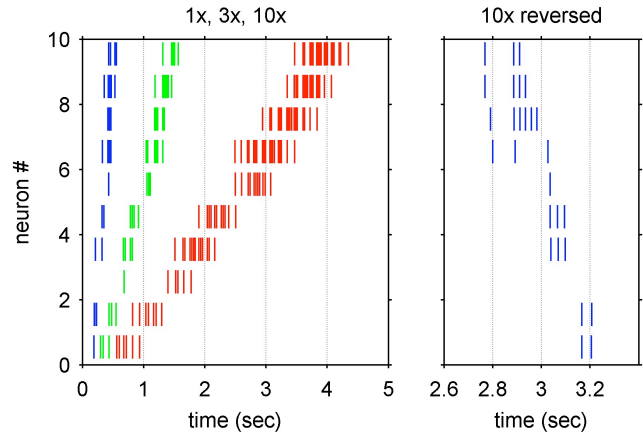

The speed of the playback from a given set of connections is determined by the rate of motion along the slow coordinate in the P neurons. For the P neuron model, the rate and direction of motion are determined by an unbalance in the steady inputs to the two sets of inhibitory cells. This allows control of the speed or even the direction of motion of the slow coordinate. For a speedup factor X, the set of S-neurons will maintain the same input currents, at a time scaled by X. This was illustrated in Fig. 1b and 1d which show that the scaling the speed of motion in the inputs from P rescales the timing of the intersecting inputs to the S neurons. The possible scalings are of course bounded: the fact that the development of synchrony in the S neurons takes a finite amount of time, and the fact that neuronal peak firing rate is also finite, are expected to limit the range of speedups at which the γ cells can respond adequately. Fig.

**Figure 5** a) Spikes from the γ cells of the designed network of Fig. 4 when the bump attractor producing the smooth P neuron firing rates moves at normal speed (red), 3x normal speed (green), and 10x normal speed (blue). b) Resulting γ spikes when the bump attractor moves in reverse at 10x normal speed.

5a shows the response of the γ cells of Fig. 4 at the original playback speed (1x), and (without changing any connections) at playback speeds of 3x and 10x. The ordering of the responses of the different γ cells is very well preserved at a speedup of 3x, and even at a speedup of 10x, the ramp-nature of the response of the γ cells can be easily distinguished. The 10x playback is reminiscent of the accelerated sequence playback observed in hippocampal cells during slow wave sleep (Lee and Wilson, 2002). Fig. 5b shows the response of the γ cells when motion in the P neurons has a reversed direction, and is at a 10-fold higher speed than the original. An orderly reversed ramp in the γ cell response can be seen. This accelerated, reversed playback of the sequence is reminiscent of the hippocampal data reported by (Foster and Wilson, 2006; Diba and Buzsáki, 2007) when rats are awake but still, and have recently navigated through a sequence of known locations.

*Single trial learning of appropriate connections*

In the circuit designed above, approximately synchronized S cells provide the excitatory drive that leads γ cells to fire. Over a brief time there is a 1:1 relationship between the S cell action potentials and the action potentials of the γ cell they are connected to. Choosing S cells with such a property can be used as the basis for learning, i.e., for selecting which S to γ cell connections should be functionally active. Let us assume that, before learning, there are no active connections between S cells and γ cells. Let us further assume that, at some given time after the P cell trajectories, there is an external sensory signal that drives a γ cell to fire at a given firing rate. For example, the γ cells could be auditory cells with frequency selectivity, cells that are activated during the hearing of a melody and that are reactivated in sequence when the melody is internally replayed. The S cells that synchronize, in a 1:1 fashion, with the γ cell at the time that the sensory input drove it to fire will be an adequate source of synaptic input with which to reproduce the γ cell's firing --at the same rate and at the same time—in the absence of the external sensory input. Using this approach with a set of γ cells, each driven by different elements of a sequence, allows an unsupervised STDP rule (Caporale and Dan, 2008) to replicate the entire sequence. We use a symmetric STDP rule similar to that of (Tsukada et al., 2005; Samura and Hattori, 2005). Below we show that the connections chosen by this neurally plausible learning method are virtually identical to the connections that would be designed to reproduce the same sequence (as done in Fig. 4).

Let a γ cell share the same theta oscillatory input and membrane time constant with the S cells. During the learning exposure, there are no synaptic inputs from S cells to γ cells. A sensory stimulus produces a brief pulse of current to the target γ cell at particular time and duration. Except for an initiation transient, the γ cell should then synchronize with a subset of S cells—the subset whose synchronized action potentials will drive the reproduction of the γ cell's firing. An STDP rule that favored creating connections between synchronized cells should thus lead to the formation of appropriate S cell to γ cell synapses. In Fig. 6, we show the action potentials of a γ cell (top, black lines) that has a membrane time constant equal to that of the S cells, and is driven by the same theta oscillation as the S cells. During the time period indicated by the grey bar, the γ cell is in addition driven by an external input current. Below it, we show the spikes of a large group of S cells, sorted by their input currents at the midpoint of the current pulse. The horizontal line at neuron 512 indicates the S cell with input current at time 2.625 sec that is closest to the input current driving the gamma cell.

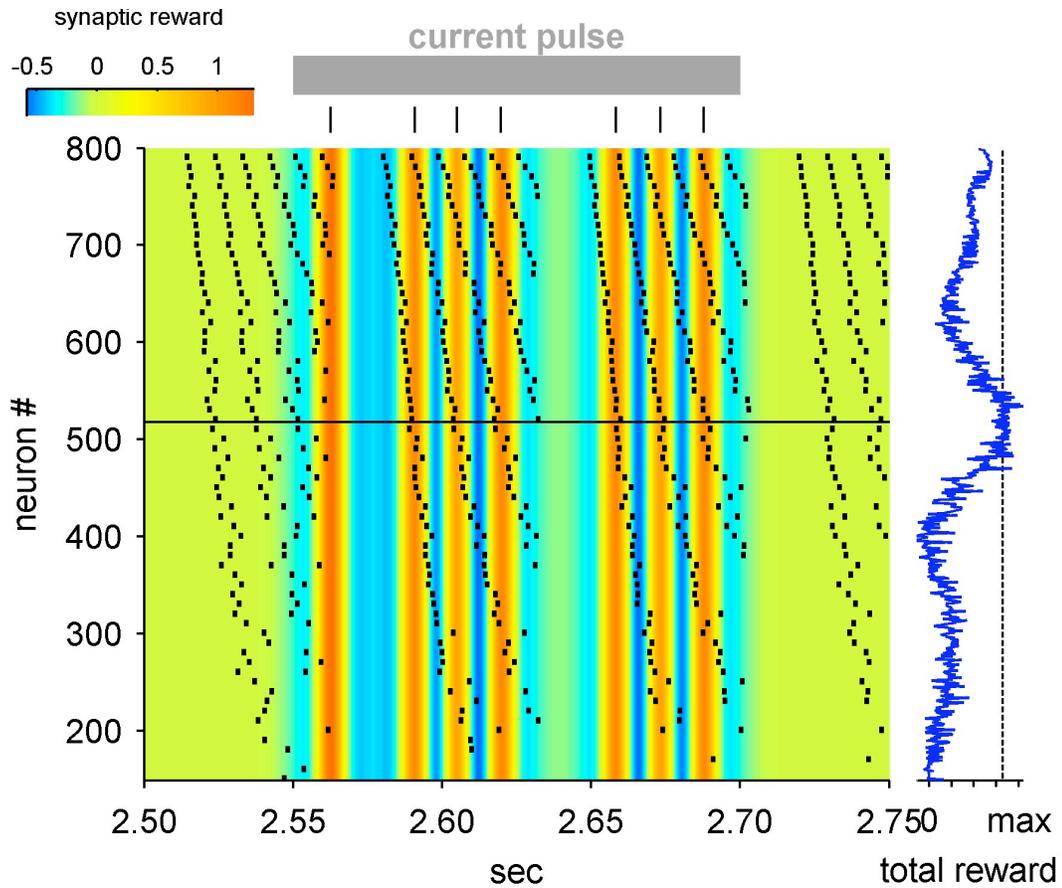

**Figure 6** Spike rasters for a set of S neurons (black square dots) and for a γ neuron (vertical lines above graph) driven by a brief current pulse centered at 2.625 sec. The S neurons are displayed in the order of their excitatory input current at 2.625 sec. The underlying color background indicates the convolution of the γ cell spikes with the STDP rule's Mexican hat function. It represents the size and sign of the STDP contribution of a presynaptic action potential occurring at any particular time, given the locations of all the post-synaptic spikes. The reward for each S cell is simply summed over the STDP value at the time of each S-cell spike, and is shown by the blue curve at the right.

We select which S cell to γ cell synapses to make active using an STDP rule with a Mexican Hat function shape and with additive contributions from different S spike/γ spike pairs. This procedure is equivalent to convolving the γ cell's spikes with the STDP rule's Mexican Hat function and then summing the values of this convolution at the location of the spikes of the S cells. The output of this procedure is what we call the "total STDP reward" for each S-cell, graphed at the right of Fig. 6. Well-synchronized S cells will have spikes mostly in the red regions, leading to a high STDP reward, while unsynchronized S cells will have spikes in both red and blue regions, leading to a low reward. As can be seen in the right panel of Fig. 6, neurons with input currents close to neuron #512 are the S neurons that are best synchronized with the cell. After exposure to a single sensory stimulus, the 30 S-cells with the greatest STDP reward are assigned connections to the γ cell, with all connections being created with equal strength. The dotted line at the right in Fig. 6 indicates the threshold value of reward for determining the connected set. In this particular example, designed connections would comprise neurons 495-525, a set very similar to the set selected by the learning rule.

The selection of S cells and consequent reproduction of the stimulus sequence is based on pattern and timing. It does not require the S cells to be identical—on the contrary, it depends on there being a diversity of S cells from which appropriate ones can be found. Our example above used S cells that differed only in their input currents merely for simplicity.

In Fig. 6 we showed a short span of time from the whole simulation, and focused on S cells that synchronized with a single γ neuron. We now take a longer stretch of time in the simulation, and consider multiple gamma neurons that can each be driven at a different time and with a different intensity. Connections from the S cell population to the gamma neurons are then learnt, for each gamma neuron, as in the single example of Fig. 6. The results of using the synaptic connections made from this learning are shown in Fig. 7.

The blue ("target") spikes in Fig. 7b are the information available to the system for learning. As with the γ cell of Fig 6, each γ cell shown in Fig. 7b is driven by a brief pulse of constant current, at a time, intensity, and duration determined by the sensory input shown in Fig. 7a. The timing of these few spikes are the sole information that the system has available for learning the desired sequence. The green spikes ("learned") in Fig. 7b show the performance of the network using the connections obtained through single trial learning. The red spikes ("designed") are reproduced here from Fig. 4 for comparison. Panels 7c,d,e use the decoding procedure described for Fig. 4 to decode each set of gamma cell spikes of Fig. 7b into a sequence. The similarity of panels 7c,d,e indicates that learning has been successful and that the learned connections are functionally equivalent to the designed connections.

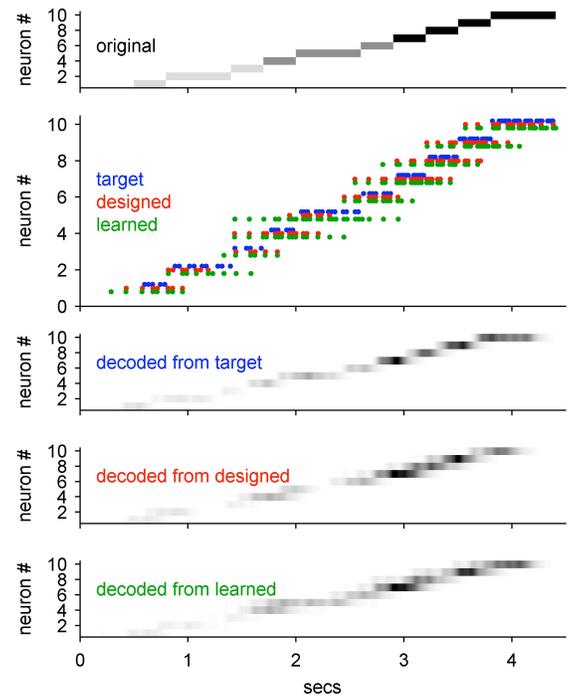

**Figure 7** Learned S to γ cell connections reproduce a target sequence of γ cell firing rates a): the intensities (gray scale) vs time for a desired sequence of 10 elements with varying intensities and durations; same data as top panel of Fig. 4. b): the action potentials of 10 γ neurons under three different conditions. "Target" spikes (blue) are spikes driven by a sequence of sensory stimuli that directly excite the γ neurons (see Fig. 1b). "Learned" spikes (green) are produced by S-> γ connections after single-trial learning. For comparison, "designed" spikes (red) from Fig. 4 are also shown. c,d,e): Corresponding decoded output intensities for the three sets of γ cell spikes from panel b (see Methods for decoding procedure).

**Discussion**

We have shown how a population of neurons with slowly varying firing rates can be used as an underlying timekeeper to drive the production of a sequence of temporal events, or to recapitulate the spatiotemporal activity pattern that has been recently experienced due to a sensory stimulus. Such slowly varying firing rates, repeatable from trial to trial, are commonly seen *in vivo* in mammals. They currently constitute a leading candidate for how time is represented in mammalian brains. While it is of course possible that the very different representation of time that has been found in songbirds also occurs in mammals, our work shows that such a representation need not be invoked—i.e., we showed that the observed slowly varying time signals are an adequate representation for such tasks.

Recognizing a particular point in time is achieved by recognizing the signature of that timepoint in the slowly-varying firing rates. This is done using the MAE ("Many Are Equals") algorithm: when Many neurons have roughly Equal firing rates, their action potentials can be synchronized, which can in turn be readily detected by a downstream γ neuron. The firing of γ neurons can encode both a motor act and its intensity. Unlike HVC-like time codes in birds, γ neurons may fire more than once in any particular sequence.

We propose that the underlying slowly varying firing rates could be generated by the slow motion of the collective state of activity of an attractor network. For the particular simulations used here, we implemented a slowly moving "bump" attractor, but other attractor networks could easily have been used. In this view, time is encoded by the position of the collective activity state. The rate of change of encoded time is therefore the same as the rate of change of the collective activity state: if this moves rapidly, sequence production can be speeded up; if the collective state moves slowly, sequence production can be slowed down; and if the collective state moves in a direction opposite to its usual motion, sequence production can be reversed. These changes in motion are readily introduced in a controlled fashion in attractor models (e.g., a leaky integrator or a drifting bump; (Zhang, 1996)). Scaling of the rate of change of slowly-varying firing rates has been observed experimentally (Komura et al., 2001; Brody et al., 2003). The speeded-up and/or reversed sequences thus produced are reminiscent of speeded-up and reversed sequences produced in hippocampal place cells during slow-wave or ripple EEG patterns, respectively (Lee and Wilson, 2002; Foster and Wilson, 2006; Diba and Buzsáki, 2007). Sequences longer than a few seconds could be achieved by linking several such slowly-moving attractor networks into a chain.

A conceptually related approach has been proposed by Matell and Meck (2004), who suggested that patterns of activity across a population of oscillating cortical neurons with diverse frequencies could be used to indicate signatures of specific points in time. If the slowly-varying input to the S neurons in our model were replaced by low-frequency sinusoids, as in the model of Matell and Meck, then longer sequences with precise timing could be learned.

We have used a new mechanism for producing the characteristic synchrony between action potentials across a chosen set of neurons that is necessary to implement MAE, namely synchronization through inhibitory conductances driven by a theta rhythm. Use of the theta-rhythm should be contrasted with previous implementations of MAE and learning. In the synaptically coupled system of *mus silicium*, (Hopfield and Brody, 2001) there were two sets of synapses that needed to be learned, and no STDP rule could be found to implement one of them. In the case of periodic gamma (Hopfield and Brody, 2004), single trial learning through STDP was effective, but the overall system had a representation of odor strength (analog value) only in a phase encoding. The theta-rhythm method we used here leads to a more easily decoded representation of intensity. In contrast, we have experimented with the noisy gamma system for the sequence production/reproduction task of the present paper, and found that in a network of the size used here multiple trials are required for robust STDP-based synaptic learning. In addition, the reproduction of intensities with the noisy gamma system was not very good for the brief intervals. Interestingly, for all four of the MAE alternatives, the desired S to γ connections are the same. We note that the sequence of events and their analog intensities are encoded in the selection of which S cells connect to which γ cells, not in the analog strength of those connections. Therefore, even when connections are learned using the theta-rhythm method, a noisy gamma signal can be added to the theta signal during playback, providing some synergistic reduction in noise. If the theta and the gamma rhythms were phase-locked, more specific synergistic effects might be expected (e.g., Jensen 2005).

We found that some additional biophysical properties, naturally occurring in neurons, were useful for improving the performance of the system. Synaptic adaptation facilitated better intensity encoding by allowing a single firing threshold in the downstream γ neurons to be appropriate for many different levels of afferent firing rates. Membrane voltage adaptation in the γ neurons allowed different cells to function effectively even if they had different numbers of synaptic inputs (a case almost unavoidable when different numbers of repeats of a given event in a sequence), again without requiring adjustments in their firing threshold.

We used an STDP rule to learn which S cells should connect to which γ cells in order to reproduce the neural activity pattern that a sensory stimulus produced. Since sequences and intensities are encoded by synapse identity, not analog strength, our synapses could be restricted to strengths of either 0 or 1. A decision between these two values can be accurately made on the basis of very few action potential pairs—few enough, in fact, that adequate learning could be obtained on the basis of a single learning trial ("one-shot" learning), with as few as four action potentials in the postsynaptic neuron. This learning process describes how we could be capable of endogenously reproducing stimulus-driven sequences of neural activities. Most experimental protocols exploring STDP use many repeated spike pairings, where learning is induced over a period of tens of seconds to minutes (Caporale and Dan, 2008). We suggest that induction of learning over a much shorter timescale – a few pairings, and hundreds of milliseconds, such as described by Rutishauser et al., 2006 – is relevant to biological tasks and merits exploration. The subjective experience of observing something only once, and rapidly (e.g., "I parked on level 3 this morning"), yet being able to remember it over timescales of many hours during which our minds are busy with other items (Standing 1970), also suggests that rapidly-induced, long-term synaptic plasticity is biologically important.

In this article, reproduction of a heard melody was used as an example of internal sequence reproduction, i.e., internally reproducing the sequence of neural activations elicited by the melody. We used examples where only single γ neurons were active at any one time, as in a melody. But the framework we have described could equally well be used to drive multiple γ neurons simultaneously, as in a sequence of chords. If the notes heard directly drive pre-motor neurons (as might be the case for someone who knows how to play the piano), then *reproduction* of this neural activation sequence in premotor cortex could be the basis for physically playing the previously heard sequence. More generally, if γ neurons are themselves motor or premotor neurons, the framework presented could be used to reproduce any desired sequence of muscle activations. Mirror neurons are an example of how observation of a sequence, produced by another, could result in an internal spatiotemporal activity pattern that represents the observed sequence if it were produced by oneself. The capacity to endogenously reproduce such an internal activity pattern, as the model presented here does, would then be the neural basis of behavioral mimicry, and the γ neurons closely related to mirror neurons.

*Predictions.* The MAE basis of our model predicts that in S neurons, which are characterized by reliable yet slowly-varying firing rates, spike time synchrony between two S neurons should depend on the magnitude of the difference between their input currents. Similar input currents should lead to greater synchrony. In many cases, firing rate can serve as a surrogate measure of input current, and therefore the prediction is that synchrony will depend on the magnitude of differences in firing rate (Markowitz 2008). If the theta rhythm is very strong, such that two neurons produce similar numbers of action potentials per theta cycle, then overall firing rate is a poor surrogate for input current strength. In this case, for theta cycles where both cells produce a spike doublet, the inter-spike interval between spikes in a doublet is a measure of input current. The prediction then is that synchrony between the *first* spike in a doublet, across the two cells, will depend on similarity, across the two cells, in the doublet inter-spike intervals.

The fact that different S neurons in the model are all reflecting the position of a single bump attractor predicts that S neurons will be correlated in a particular way. Let us assume that, over different productions of the same sequence, the speed of P neuron firing rate evolution may vary slightly, in some trials being faster than others. All S neurons will follow the same speed-up/slow-down pattern. Thus, single-neuron per-trial estimates of overall speed-up or slow-down should be correlated across different of neurons.

**Methods**

*Computation*

All simulations were carried out in Matlab. The differential equations representing neural variables were integrated using an Euler procedure with a 0.1 ms time step. Simulation code is available at http://genomics.princeton.edu/hopfield/simulation.html

*Attractor network for generating slowly varying inputs to S neurons*

A biological example of the slowly changing signals on which timing is to be based is illustrated in Fig. 1a. Such patterns are characteristic of a class of attractor network. If for the moment we consider simple rate-based neural models, then the present state of the system can be specified by the firing rates of the N cells, and is represented by a point in N-space. If the nature of the connectivity results in the evolution of the activity in time proceeding until the activity lies on a particular 1-dimensional curve in this N-space and then stops, the system is described as having a line attractor. The neural integrator in the vestibular-ocular system has been described as a line attractor, as has the bump attractor in the head-direction remembering system. The neural dynamics of these systems is described by N-1 dimensions in which the system rapidly evolves, and one dimension along which it does not change with time. The behavior illustrated in Fig 1a is a minor variant of the line attractor in which there is rapid motion in the activity until the state lies on a 1-d curve in N-space, and then a slow, typically linear, drifting motion along that curve. The speed of drift is controlled by a single input parameter. The activity of each neuron will then change slowly in time, and the coordination between these activities is fixed by the connectivity pattern. A 'leaky integrator' or 'ring attractor network with drift' are particular examples of such systems. We have chosen to model this part of the system, the P cell box of Fig 1, as section cut from a bump attractor network in close correspondence to an established model (Song and Wang, 2005), except that spiking neurons were replace by rate-based units for simplicity and speed of computation. The model was chosen for biological plausibility (unlike many bump attractor models it obeys Dale's law) and for the fact that the rate of drift along the slow dimension was controlled by plausible inputs. Excitatory currents proportional to these firing rates are used to drive the cells of box 4 in the model of Fig. 1d. The stereotyped time responses seen in *vivo* are initiated by a 'start' signal through some unknown neural process, which can be either internally generated ('recall') or externally generated ('listening and remembering') To achieve the same end, we directly initiate the 'drifting attractor network' at the start of either task by briefly initiating a bump of fixed location in the bump attractor. Following this initiation, the leaky integrator follows its slow dynamics along its stereotype 1-dimensional curve, producing the time-dependent signals illustrated on the right side of Fig. 1. It should be emphasized that for the general phenomena to be described in this paper, the origin of these time varying signals is unimportant. It is important only that they are coordinated along a slow drift coordinate, and have the slow rate of change and diversity of behaviors like those observed in frontal cortex.

The beginning of any sequence initializes the bump attractor, and therefore the P neuron firing rates, to a standard state, which then evolves slowly in time. Both the initial state and the time-evolution of P neurons are independent of the stimulus or the particular sequence to be played back. The MAE approach benefits from a large diversity of neurons that have similar firing rates at any given point in time. To provide this diversity, different S neurons in the model were driven with different input strengths from P neurons in the bump attractor. Each S neuron was driven by one P bump attractor neuron.

*S cells*

These cells are modeled as leaky integrate-and-fire neurons with a membrane time constant of 15 ms. Spike generation is instantaneous when the membrane potential reaches a threshold value of 20 mV, and the cell potential is then reset to zero. After a spike, the membrane potential is held at zero for 2 ms, representing an absolute refractory period. The resting potential of these neurons is 9 mV. An

In most of the simulations, all S neurons receive a common inhibitory synaptic input with a theta-rhythm. This input is modeled as though it came from synchronous volleys of action potentials from an inhibitory network, activating the conductivity of channels for an ion with a reversal potential of 0 mV. These pulses occur in a quasi-random fashion, with intervals between them randomly and uniformly chosen in the interval 100 -160 ms (6.7 Hz to 10 Hz). This randomness was introduced to emphasize the fact that for this means of evaluating MAE, the underlying rhythm need not be periodic. The conductivity pulses were modeled as alpha-functions with a characteristic time of 10ms. The peak conductivity due to this inhibitory synapse was equal to the conductivity of the cell membrane in the absence of input. The timing of the initiation of the slow input bump attractor was independent of the theta pulses.

The excitatory synapses from S neurons to γ neurons were modeled as the conductivity of AMPA-like receptors with time dependence

$$\sigma_{synaptic} = (1 - \exp(-t/\tau_{on}))*(\exp(-t/\tau_{off}))$$

with $\tau_{on}$ = 0.7 ms and $\tau_{off}$ = 1.5 ms. The reversal potential for $Na^+$ was 100 mV. All non-zero synapses had the same strength.

Rate-dependent adaptation for S neurons to γ neurons synapses was modeled as follows. After each synaptic conductivity pulse due to a presynaptic action potential, the synaptic efficacy is reduced by a factor (1 – 0.07). (In biological synapses, vesicle depletion in a pool of 15 vesicles could have such an effect.) This reduction disappears with a time constant of 400 ms. For steady state synaptic use at firing rate f, the synaptic efficacy as a function of firing rate in this model would be 1/( 1 + 0.028*f). This weak adaptive behavior allows gamma neurons to have the same firing threshold and maintain their stimulus selectivity, even while they may be driven at quite different firing rates (Tsodyks and Markram, 1997).

### γ cells

There are two phases to the use of γ cells, the learning phase and the performance phase. During the learning phase, a particular γ cell is driven by a current pulse coming from the sensory system some time after the initiation cue. Which cell gets driven when depends on the encoding between the sensory system and the γ cells. For simple melody encoding, each γ cell corresponds to a particular musical note. In the learning phase, each γ cell is driven at a time that describes when the note is to occur with respect to the initiation cue. The strength of the driving current pulse describes the intensity of the note.

In the learning phase, γ cell electrical characteristics are identical to the S cells, and the γ cells receive the same inhibitory theta pulses as the S cells. In the learning phase, there are silent synapses between each S cells and each γ cell.

In the performance phase there are three changes to the γ cell behavior. None is necessary to obtain crude operation, but all help to obtain good performance. First, the time constant of the γ cells is shortened from 15 ms to 5 ms. Second, the theta rhythm inhibitory input is turned off. Third, an adapting voltage-dependent 'conductivity channel' is introduced for simulations late in the paper where repeats are involved (Figs. 8,9). These channels make it more difficult to drive the cells to firing if they have had elevated membrane potentials recently (*i.e.*, post depolarization depression), and easier to make the cell fire if the membrane potential has been held at low levels (post-hyperpolarization excitability). When found in biological neurons, such a behavior is typically described in terms of adaptive gain rescaling. It plays the same role here in modifying the excitability of the γ neurons on the basis of the level of ongoing synaptic input currents, even when that input is not large enough to make the γ neuron fire.

The adaptation was modeled as follows. A variable u_bar, the recent time-average of the membrane potential, was computed according to

$$du\_bar/dt = -(u\_bar - u)/\tau_{adapt}$$

with a constraint that keeps u_bar >= 0. Since u_bar is positive, it can be understood conceptually as a stand-in for the details of the voltage-dependent site occupancy of a channel ligand. $\tau_{adapt}$ was not a sensitive parameter, and values in the range 100-250 ms give similar results. This channel controlled an inhibitory conductivity channel of an ion having a reversal potential of 0 mV according to

$$\sigma_{adapt} = (u\_bar - u\_bar_{threshold}) \text{ for } u\_bar > u\_bar_{threshold}, \text{ and zero otherwise.}$$

u_bar$_{threshold}$ = 4 mV.

This results in behavior similar to that found by Spain and colleagues (Spain et al., 1991). During the simulations the value of u_bar lies in the range 4 < u_bar <10 and the non-negativity constraints are inoperative. They have been included in the simulation code only because non-negativity is required of a reasonable biophysical model, and because others may wish to use the code in a more extended range

*'Neural' decode of spike trains into analog values*

There are many ways to 'neurally' decode such action potential trains into analog signals. We illustrate (and have used) a particular one, to show by example that if in biology there were a need to produce an analog signal that resembles the actual input analog intensities, such a signal can be generated even from the properties of simple neurons. While this procedure is based on cell and membrane properties, other procedures involving synaptic properties such as facilitation or metabatropic action are readily constructed.

The action potentials from each gamma cell were used to drive 1000 identical integrate-and-fire decoding neurons through excitatory synapses. The intrinsic properties of these neurons are exactly the same as the S-cells. The resting potential was set 1.6 mV below the threshold for firing. The random current injected to simulate membrane noise was increased by 60% to decrease the number of cells necessary to get good statistical smoothing from the noise. The excitatory synaptic currents to the decoding neurons are slow; an action potential from a gamma neuron produces an input current to the decoding neuron soma having a shape $(t/\tau_{syn})^2 \exp(-t/\tau_{syn})$ with $\tau_{syn}$ = 30 ms. While this does not correspond to any simple channel in a compact neuron, such a slowed form can biologically result from dendritic integration and delay of fast excitatory currents in a non-compact cell, since the $\tau_{syn}$ used in reconstruction is similar to the membrane time constant. The spike rasters of these 1000 decoding neurons are summed. This action potential sum was smoothed for use in the figures with an $\alpha$-function smoothing kernel with time constant 60 ms.

*learning protocol*

Learning is particularly simple to do when the connections to each γ neuron need to represent only a single event during the entire time interval. The γ neuron generates spikes only during the single event. Spike timings between those spikes and the spiking of each S neuron can be used with a spike-timing-dependent protocol for choosing which of the silent synapses to convert to AMPA-like synapses.

A 'reward' was calculated for each synapse, summing the contribution of all pre-post synaptic spike pairs separated by a time difference δt milliseconds according to

rewardshape = $\exp(-(\delta t/10)^2) - 0.35*\exp(-(\delta t/25)^2)$

After a single experience of the pattern to be learned, the 30 synapses with the greatest reward are made into active synapses, all with the same strength.

*learning when some γ cells need to respond more than once*

If we wish to make a γ cell respond at two different times, and with an appropriate representation of intensity at both times, a designed solution to this task is to simply make two sets of synapses onto the cell, one corresponding to each time and intensity. At each recognition epoch, the γ cell receives synaptic input from both. The presynaptic action potentials for one are well-synchronized; the presynaptic action potentials for the other are disorganized, and to a first approximation provide only a DC level shift. This level shift is removed by the slow adaptive nature of the γ cells. As a result, this set of designed connections functions well. Our remaining task is to identify a STD learning protocol that will generate these connections.

We initially tried to merely use the procedure defined in the previous section by simply increasing the number of synapses chosen.  This procedure seldom gave good results because it lacked essential balance.   If one epoch to be remembered of greater duration, or more intense, the synapse selection procedure would choose concentrate too many synapses on this epoch and two few on the other.  In addition, S cells are chosen that will somewhat contribute at both times, rather than to be good for one time or for the other.  We believe that any procedure that waits for the end of the entire sequence to learn, sums over all experienced pre-post spike timing pairs at each synapse, and then chooses synapses, will have this failing.  Instead, what is needed is an automatic segmentation into learning epochs in time.  When the end of a segment is identified, a synapse change algorithm based on all the spike pair timings accumulated in that epoch results, and accumulation of further spike-pair timings for the next epoch is initiated.

During the learning phase, the γ cell has no synaptic drive from S cells, and is driven only by the sensory input that consists of two unequal pulses separated in time.  As a result, the γ neuron produces the spikes shown in Fig. 9b).   Any procedure that automatically identifies times shortly after bursts of spikes as the ends of learning epochs will suffice to break the learning into two unrelated segments as desired.  Because spike pairs with large time differences do not contribute to STDP rules, results are insensitive to exactly where these end-of-epoch fiduciaries are drawn.   We modeled this process by introducing a variable C mimicking a $Ca^{++}$ high potential conductivity channel and a $Ca^{++}$ extrusion system.  This variable was incremented up a fixed amount every time the γ cell produced an action potential, and decayed exponentially with a time constant of 100 ms.  When a γ cell has a single burst of activity, C displays a single broad peak near the time of that burst, and is in a falling phase at the conclusion of the burst.  A fiducial time after that burst can be defined in simple Matlab code by identifying the location where C has a downward crossing of a threshold level.  To make such simple code Matlab code work effectively, some jaggedness in C due to the discrete action potentials was removed by an additional 100 ms smoothing filter.  When the γ cell produces two bursts of spikes, two downward crossings occur.  These crossings identified the times at which to learn, providing automatic segmentation of the learning process described in the previous paragraph.

**Acknowledgments** We thank David Tank, Forrest Collman and David Markowitz for comments on the manuscript, and we thank Ranulfo Romo for kindly sharing his prefrontal cortex data.

# Supplementary Information

*Learning with repeated elements in a sequence*

So far we have dealt with only sequences having no repeated element.  Can a network be constructed to learn the connections when the same γ neuron is to be active at two different times in the sequence?  A possible design for an appropriate set of connections is to use twice as many synaptic connections of the same strength for that gamma neuron, comprised of the two different sets of connections that would have been previously used for sequences with no repeats.  This simple approach does not quite work.  Since all inputs to the γ cell are excitatory, when the number of input synapses is doubled, even partial synchronization drives the γ cell to fire. This results in the γ cell firing most of the time instead of selectively firing at two appropriate time epochs.  The problem is seen in Fig 8a), where the membrane potential of γ cells (which are being prevented from generating action potentials) with one set of connections and with two sets of connections are shown.  No single threshold will both generate two clumps of spikes driven by the repeated occurrence system and also produce reliable spikes from the single occurrence system.

If the number of synapses is to be doubled at the given fixed strength, some kind of adaptation in the gamma cell is necessary to compensate. This adaptation must take place when the γ cell is *not* firing, and thus be based on sub-threshold membrane properties. The known phenomenon of post-hyperpolarization excitation (Spain et al., 1991) forms the basis for such adaptation. In this phenomenon, neurons that have been held more hyperpolarized are more readily driven by an excitatory current compared to neurons that have been held less hyperpolarized. Within this paradigm, neurons with more synapses to them will have bigger input currents to them when they are not firing, will have less depolarized membrane potentials, and will be harder to drive to fire when compared to neurons that have less synapses to them.

The mathematical model we have chosen for post-hyperpolarization excitation is based on a voltage-dependent inhibitory channel conductivity that has slow kinetics that average over a time $\tau_{adapt}$. A channel variable $u_{bar}$ obeys

$$du_{bar}/dt = (u_{bar} - u)/\tau_{adapt}$$

and controls the channel conductivity through the relationship

$\sigma_{channel} = 0$ for $u_{bar} < 4$ mV        $\sigma_{channel} = \sigma_0 (u_{bar} - 4)$ for $u_{bar} > 4$ mV

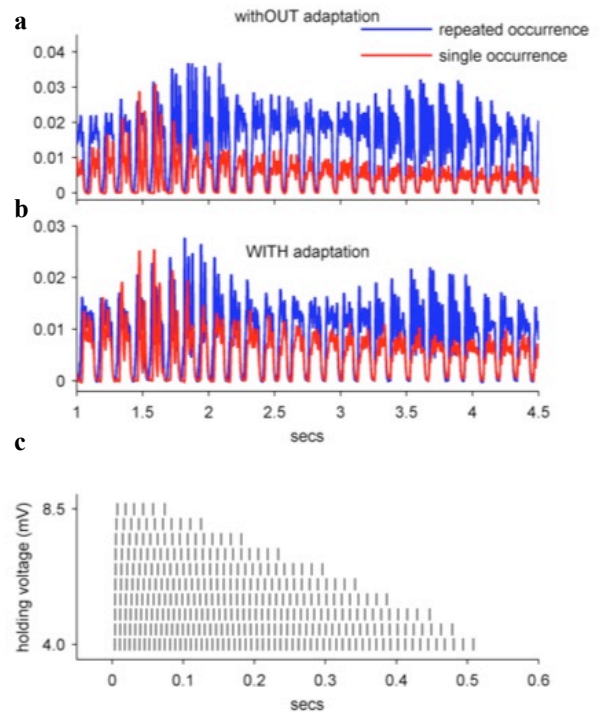

**Figure 8** The membrane potentials for two γ cells with action potentials suppressed without a) and with b) cellular adaptation. c) Each row raster is the result of holding a γ cell in voltage clamp at the holding voltage indicated on the y-axis, below the firing threshold of 10 mV, and moving it to current clamp at t = 0. The clamped current is the same for each raster.

The ion for which the channel conducts was taken to have a Nernst potential of zero. This conductivity is in parallel with the intrinsic membrane conductivity, which was somewhat reduced from previous values to make the net time constant reasonable. The effect of this adaptation is shown in Fig. 8b). With this adaptation it is now possible to set a single threshold that will yield appropriate spikes for both gamma cells.

These γ cells now exhibit post-hyperpolarization excitation. Fig. 8c) shows the action potentials that such a cell produces when the membrane voltage u was clamped at different holding voltages before t = 0 and switched to current clamp for t > 0 with the current the same for all holding voltages. The lower the holding voltage, the higher the initial rate of firing and the longer the firing continues.

With adapting gamma neurons, the designed connections successfully reproduce sequences having recurring elements. This is illustrated in Fig. 9, which is closely parallel to Fig 4 except that the sequence now has some repeated elements and the gamma cells are adapting. With this adaptation, the γ cell effective time constant is about 5 ms for the case of a repeated occurrence, and 8 ms for the single occurrence.

Learning a set of connections that can deal with repeated elements requires identifying independent learning epochs, and making the synapses that are appropriate to each epoch. An approach that attempts to first sum 'synaptic reward' over all time, and then make a decision about what synapses to make on the basis of the size of that sum, is conceptually flawed. It will result in a synaptic imbalance between the two epochs if the epochs differ in either duration or intensity. In addition, such an approach will choose synapses that are somewhat in support of both epochs in preference to those which are truly appropriate to either.

What is needed is a learning protocol that identifies when a γ cell has recently been quite active and now has gone silent, indicating that now is the time to learn an epoch. The synaptic change at that time should be based on a short-term average of the reward based on recent pre-post synaptic firing, making excitatory synapses of uniform strength at the synapses with greatest accumulated reward. At that learning time, the reward average should be reset to zero in preparation for another learning epoch. Any implementation of these ideas will succeed in choosing the correct synapses for multiple occurrences of an element. In the simulations shown below, we assumed $Ca^{++}$ influx on gamma cell spiking (see Methods). A downwards crossing through a high $Ca^{++}$ concentration threshold was then used to identify moments when the gamma cell had been spiking and now fell silent. The synaptic reward was computed from the spike-timing rules of Fig. 4 and averaged over the period since the last reset of the reward integrator (either the start of the trial or the last $Ca^{++}$ downwards-threshold). We emphasize that the specifics of how this was done are not important for the final results. The results of such a procedure are shown in Fig. 9b and 9e. The γ cell spikes that result from these learned connections, and the decoded sequence they produce, are similar to the spikes and sequence produced by the designed connections.

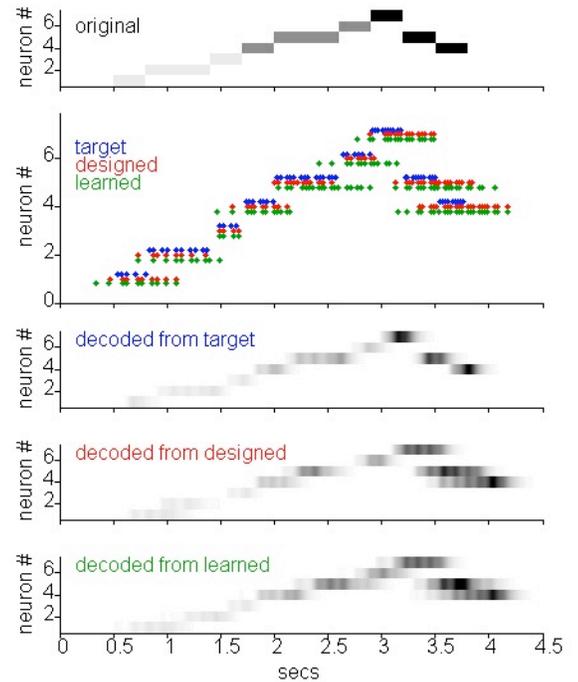

**Figure 9.** Comparison of the γ cell spikes and reconstructions of a sequence for a sequence with repeats in elements 4 and 5. a) The sequence to be reproduced, as in Fig. 4 and 7. b) The γ cell spikes for the target sequence and for designed and learned connections. c-e) The reconstructed sequences from the three sets of γ cell spikes.

## References

Abbott, L.F., Varela, J.A., Sen, K., and Nelson, S.B. (1997). Synaptic depression and cortical gain control. Science *275*, 220-224.

Brody, C.D., and Hopfield, J.J. (2003). Simple networks for spike-timing-based computation, with application to olfactory processing. Neuron *37*, 843-852.

Brody, C.D., Hernández, A., Zainos, A., and Romo, R. (2003). Timing and neural encoding of somatosensory parametric working memory in macaque prefrontal cortex. Cereb Cortex *13*, 1196-1207.

Brozović, M., Gail, A., and Andersen, R.A. (2007). Gain mechanisms for contextually guided visuomotor transformations. J Neurosci *27*, 10588-10596.

Catania, A.C. (1970). Reinforcement schedules and psycholphysical judgements. In The theory of reinforcement schedules, W. N. Schoenfeld eds., New York, pp. 1-42

Caporale, N., and Dan, Y. (2008). Spike timing-dependent plasticity: a Hebbian learning rule. Annu Rev Neurosci *31*, 25-46.

Diba, K., and Buzsáki, G. (2007). Forward and reverse hippocampal place-cell sequences during ripples. Nat Neurosci *10*, 1241-1242.

Foster, D.J., and Wilson, M.A. (2006). Reverse replay of behavioural sequences in hippocampal place cells during the awake state. Nature *440*, 680-683.

Hahnloser, R.H., Kozhevnikov, A.A., and Fee, M.S. (2002). An ultra-sparse code underlies the generation of neural sequences in a songbird. Nature *419*, 65-70.

Hopfield, J.J., and Brody, C.D. (2000). What is a moment? "Cortical" sensory integration over a brief interval. Proc Natl Acad Sci U S A *97*, 13919-13924.

Hopfield, J.J., and Brody, C.D. (2001). What is a moment? Transient synchrony as a collective mechanism for spatiotemporal integration. Proc Natl Acad Sci U S A *98*, 1282-1287.

Hopfield, J.J., and Brody, C.D. (2004). Learning rules and network repair in spike-timing-based computation networks. Proc Natl Acad Sci U S A *101*, 337-342.

Jensen, O., and Lisman, J.E. (2005). Hippocampal sequence encoding driven by a cortical multi-item working memory buffer. Trends Neurosci. 28, 67-72

Kojima, S., and Goldman-Rakic, P.S. (1982). Delay-related activity of prefrontal neurons in rhesus monkeys performing delayed response. Brain Res *248*, 43-49.

Komura, Y., Tamura, R., Uwano, T., Nishijo, H., Kaga, K., and Ono, T. (2001). Retrospective and prospective coding for predicted reward in the sensory thalamus. Nature *412*, 546-549.

Lee, A.K., and Wilson, M.A. (2002). Memory of sequential experience in the hippocampus during slow wave sleep. Neuron *36*, 1183-1194.

Levitan, W., and Kaczmarek (1996). The Neuron: Cell and Molecular Biology (Oxford: Oxford University Press).

Markowitz, D.A., Collman, F., Brody, C.D., Hopfield, J.J., and Tank, D.W. (2008). Rate-specific synchrony: using noisy oscillations to detect equally active neurons. Proc Natl Acad Sci U S A *105*, 8422-8427.

Rutishauser, U., Mamelak, A.N., Schuman, E.M. (2006). Single-trial learning of novel stimuli by individual neurons of the human hippocampus-amygdala complex. Neuron, *49*, 805-813.

Matell, M. S. and W. H. Meck (2004). Cortico-striatal circuits and interval timing: coincidence detection of oscillatory processes. Cognitive Brain Research *21*: 139-170.

Mauk, M.D., and Buonomano, D.V. (2004). The neural basis of temporal processing. Annu Rev Neurosci *27*, 307-340.

Meck, W.H. (2005). Neuropsychology of timing and time perception. Brain Cogn *58*, 1-8.

Mita, A., Mushiake H., Shima K., Matsuzaka Y., and Tanji, J. (2009). Interval time coding by neurons in the presupplementary and supplementary motor areas. Nat. Neurosci., *12,* 502-507.

Roberts, S. (1981) Isolation of an internal clock. J. Exp. Psychol., Anim. Behav. Process. *7*:242-268

Samura, T., and Hattori, M. (2005). Hippocampal memory modification induced by pattern completion and spike-timing dependent synaptic plasticity. Int J Neural Syst *15*, 13-22.

Song, P., and Wang, X.J. (2005). Angular path integration by moving "hill of activity": a spiking neuron model without recurrent excitation of the head-direction system. J Neurosci *25*, 1002-1014.

Spain, W.J., Schwindt, P.C., and Crill, W.E. (1991). Two transient potassium currents in layer V pyramidal neurones from cat sensorimotor cortex. J Physiol *434*, 591-607.

Standing, L., Conizio, J., and Haber, R. N. (1970). Perception and memory for pictures: single-trial learning of 2,500 individual stimuli. Psychon. Sci. *19*, 73-74

Tsodyks, M.V., and Markram, H. (1997). The neural code between neocortical pyramidal neurons depends on neurotransmitter release probability. Proc Natl Acad Sci U S A *94*, 719-723.

Tsukada, M., Aihara, T., Kobayashi, Y., and Shimazaki, H. (2005). Spatial analysis of spike-timing-dependent LTP and LTD in the CA1 area of hippocampal slices using optical imaging. Hippocampus *15*, 104-109.

Wang, X.J. (2001). Synaptic reverberation underlying mnemonic persistent activity. Trends Neurosci *24*, 455-463.

Zhang, K. (1996). Representation of spatial orientation by the intrinsic dynamics of the head-direction cell ensemble: a theory. J Neurosci *16*, 2112-2126.

Zipser, D., and Andersen, R.A. (1988). A back-propagation programmed network that simulates response properties of a subset of posterior parietal neurons. Nature *331*, 679-684.